\documentclass[fleqn,twoside]{article}
\usepackage{espcrc2}
\usepackage{url}

\usepackage{graphicx}
\usepackage[figuresright]{rotating}

\newcommand{\AmS}{{\protect\the\textfont2
  A\kern-.1667em\lower.5ex\hbox{M}\kern-.125emS}}

\title{Pion Form Factor at SND ({\it new edition}). %\Latex
}

\author{M.N. Achasov \address[BINP]{Budker Institute of Nuclear Physics,
                                    Siberian Branch of the Russian Academy of 
				    Sciences, \\ 11 Lavrentyev,Novosibirsk,
				    630090, Russia}%
 \thanks{E-mail:achasov@inp.nsk.su.},
 K.I.Beloborodov\addressmark[BINP]\address[NSU]{Novosibirsk State University, 
                                                630090, Novosibirsk, Russia},
 A.V.Berdyugin\addressmark[BINP],
 A.G.Bogdanchikov\addressmark[BINP],
 A.V.Bozhenok\addressmark[BINP]\addressmark[NSU],
 A.D.Bukin\addressmark[BINP],
 D.A.Bukin\addressmark[BINP],
 T.V.Dimova\addressmark[BINP],
 V.P.Druzhinin\addressmark[BINP]\addressmark[NSU],
 V.B.Golubev\addressmark[BINP]\addressmark[NSU],
 A.A.Korol\addressmark[BINP],
 S.V.Koshuba\addressmark[BINP],
 E.V.Pakhtusova\addressmark[BINP],
 S.I.Serednyakov\addressmark[BINP]\addressmark[NSU],
 Yu.M.Shatunov\addressmark[BINP]\addressmark[NSU],
 V.A.Sidorov\addressmark[BINP],
 Z.K.Silagadze\addressmark[BINP]\addressmark[NSU],
 A.N.Skrinsky\addressmark[BINP],
 Yu.A.Tikhonov\addressmark[BINP]\addressmark[NSU], and
 A.V.Vasiljev\addressmark[BINP]\addressmark[NSU],}
       
\begin{document}

\begin{abstract}
 The update of the $e^+e^-\to\pi^+\pi^-$ process cross section, measured in 
 the energy region $\sqrt{s}<1$ GeV with SND detector at VEPP-2M collider
 is presented.
 
\vspace{1pc}
\end{abstract}

\maketitle

 The $e^+e^-\to\pi^+\pi^-$ process cross section was measured by SND 
 \cite{sndnim} detector at VEPP-2M \cite{vepp2} collider in the energy region
 below 1 GeV  \cite{snd-2pi}. The measurement was based on about 
 $5\times 10^6$ $e^+e^-\to\pi^+\pi^-$ events and has systematic accuracy 1.3\%.

\begin{figure}[h]
\includegraphics[scale=0.5]{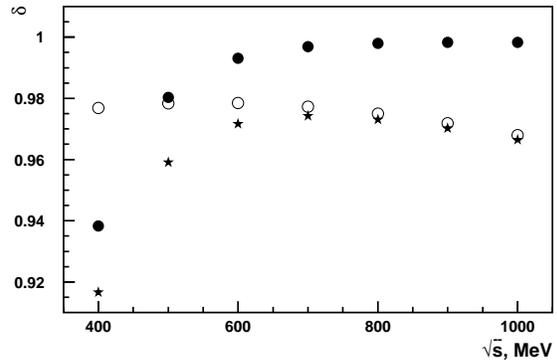}
\caption{Corrections $\delta$ to the $e^+e^-\to\pi^+\pi^-$ cross section.
         $\bullet$ and $\circ$ are corrections due to inaccuracy of the
	 $e^+e^-\to\mu^+\mu^-\gamma$ and $\pi^+\pi^-\gamma$ events generator,
	 $\star$ is the total correction.}
\label{nonp}
\end{figure}

 Calculation of the $e^+e^-\to e^+e^-\gamma$, $\pi^+\pi^-\gamma$ and
 $\mu^+\mu^-\gamma$ processes cross sections plays quite an important role in
 form factor measurement. It is necessary for the luminosity measurement 
 ($e^+e^-\to e^+e^-\gamma$), for background subtraction
 ($e^+e^-\to\mu^+\mu^-\gamma$), for radiative corrections and detection
 efficiency determination ($e^+e^-\to\pi^+\pi^-\gamma$). Recently it was
 found, that $e^+e^-\to\pi^+\pi^-\gamma$ and $e^+e^-\to\mu^+\mu^-\gamma$
 Monte Carlo events generators based on Ref.\cite{gpp,gmm} were not quite
 correct. So we performed reanalysis of SND data using MCGPJ generator
 \cite{mcgpj}, which is based on the same articles. 
 As a result of reanalysis
 the corrections to the measured $e^+e^-\to\pi^+\pi^-$ cross section 
 \cite{snd-2pi} were obtained (fig.\ref{nonp}). After the corrections the 
 cross section values decreased by two systematic errors.

\begin{figure}
\includegraphics[scale=0.5]{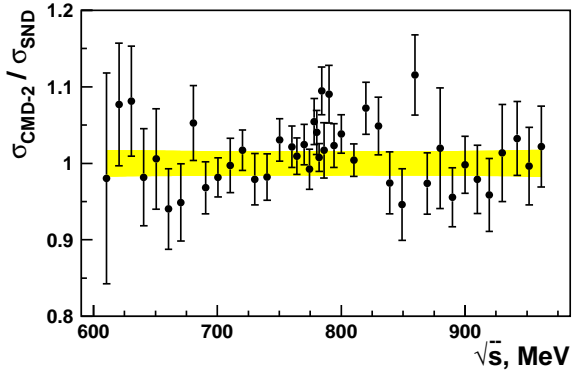}
\caption{The ratio of the $e^+e^-\to\pi^+\pi^-$ cross section obtained in CMD-2
         \cite{2pi-cmd-1} and SND (this talk) experiments. The shaded area
         shows the joint systematic error.}
\label{cp1}
\end{figure}

\begin{figure}[h]
\includegraphics[scale=0.5]{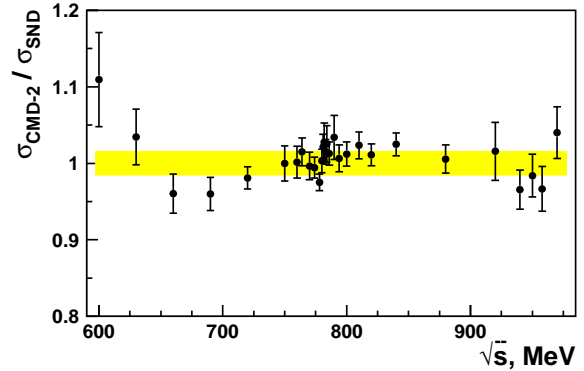}
\caption{The ratio of the $e^+e^-\to\pi^+\pi^-$ cross section obtained in CMD-2
         \cite{2pi-cmd-2} and SND (this talk) experiments. The shaded area
         shows the joint systematic error.}
\label{cp2}
\end{figure}

 The comparison of corrected SND result with the CMD-2 precision  
 measurements \cite{2pi-cmd-1,2pi-cmd-2} is shown in  fig.\ref{cp1} and 
 \ref{cp2}. 
 In the energy region $\sqrt{s}$ from 600 to 1000 MeV the average deviation
 \begin{equation}
 \Delta = (1 - \langle {\sigma_{CMD-2} / \sigma_{SND}}\rangle )\times 100\%  
\end{equation}
 between the SND ($\sigma_{SND}$) and CMD-2 ($\sigma_{CMD-2}$) measurements is
 equal to $-1.4\pm 1.5 \%$ for the published \cite{2pi-cmd-1} CMD-2 data and 
 $-0.3\pm 1.6 \%$ for the new, not yet published but already reported, CMD-2 
 result \cite{2pi-cmd-2}. For uncorrected SND cross section $\Delta$ was equal
 to $1.4\pm 1.5 \%$ \cite{snd-2pi} and $2.5\pm 1.6 \%$ \cite{itep05}
 respectively. The  errors include both systematic and statistic uncertainties 
 of the CMD-2 and SND measurements.

 The table of corrected cross sections and results of theoretical fit to the
 data will be published soon.

 The authors are grateful to G.V. Fedotovich, F.V. Ignatov, A.L. Sibidanov
 for useful discussions.
 The work is supported in part by grants Sci.School-1335.2003.2,
 RFBR 04-02-16181-a, 04-02-16184-a, 05-02-16250-a, 06-02-16192-a.


\begin{thebibliography}{9}
\bibitem{vepp2}
 A.N. Skrinsky, in Proceedings of Workshop on physics and detectors for
 DA$\Phi$NE, Frascati, Italy, April 4-7, 1995, p.3.
\bibitem{sndnim}
 M.N. Achasov et al., Nucl. Instr. and Meth. A 449 (2000) 125
\bibitem{snd-2pi} M.N. Achasov et al., Zh. Eksp. Teor. Fiz. B 128 (2005) 1201;
 JETP 101 (2005) 1053.
\bibitem{gmm} A.B. Arbuzov et al., JHEP 10 (1997) 001
\bibitem{gpp} A.B. Arbuzov et al., JHEP 10 (1997) 006 
\bibitem{mcgpj} A.B. Arbuzov et al., Report of Budker INP 2004-70, 
                Novosibirsk (2004); hep-ph/0504233;
 Computer code was got from \url{http://cmd.inp.nsk.su/~sibid/radnsk.tar.bz2}
\bibitem{2pi-cmd-1}
  R.R. Akhmetshin et al, Phys. Lett. B 527 (2002) 161;
  Phys. Lett. B 578 (2004) 285  
\bibitem{2pi-cmd-2}
 I.B. Logashenko et al., in Proceedings of International European Conference
 on High Energy Physics, Lisboa, Portugal, July 21-27, 2005; 
 A.L. Sibidanov et al., in Proceedings of 11 International Conference on
 Hadron Spectroscopy, Rio-De-Janeiro, Brazil, August 21-26, 2005;
 I.B. Logashenko, this Proceedings
\bibitem{itep05}
 M.N. Achasov, in Proceedings of Conf. ``Physics of fundamental interactions'',
 dedicated to 60th anniversary of ITEP (Session of RAS Nuclear Physics 
 Department 2005), Moscow, December 5-9, 2005
\end{thebibliography}
\end{document}